\documentclass[aps, showpacs, prl, superscriptadaddress, %eqsecnum, 
twocolumn] {revtex4-1}
\usepackage{hyperref}
\usepackage{amsmath }
\usepackage{amssymb}
\usepackage{epsfig}
\usepackage{natbib}
\usepackage{epstopdf}
\newcommand*{\be}{\begin{equation}}
\newcommand*{\ee}{\end{equation}}
\newcommand*{\bea}{\begin{eqnarray}}
\newcommand*{\eea}{\end{eqnarray}}
 
 \DeclareFontFamily{OT1}{pzc}{}
 \DeclareFontShape{OT1}{pzc}{m}{it}%
 {<->  s  *  [1.400]  pzcmi7t}{}
\DeclareMathAlphabet{\mathscr}{OT1}{pzc}%
{m}{it}

\begin{document}

%\title[]{Optical coupler in active medium}
\title{Actively coupled optical waveguides}

\author{N. V.  Alexeeva$^{1,2}$, I. V. Barashenkov$^{1,2}$,  K. Rayanov$^1$ and S. Flach$^1$  }
 \affiliation{%$^2$
  New Zealand Institute for Advanced Study, 
Centre for Theoretical Chemistry and Physics, Massey University, Auckland 0745, New Zealand 
\\  $^2$  
Department of Mathematics and Centre for Theoretical  and Mathematical Physics,  University of Cape Town, Rondebosch 7701, South Africa 
 }

\begin{abstract}
We consider light propagation through
 a pair of nonlinear optical waveguides with absorption, placed in a medium with power gain.
 The active medium boosts the in-phase component of the 
  overlapping evanescent fields of the guides,
 while the nonlinearity of the guides couples  it to the damped
 out-of-phase component
 creating a feedback loop. 
  %When this active coupling exceeds a threshold in the presence of Kerr nonlinearity,
As a result, the structure exhibits stable stationary and oscillatory regimes in a wide range of gain-loss ratios.
We show that the pair of actively-coupled  ($\mathcal{AC}$) waveguides can act as a 
stationary or integrate-and-fire comparator sensitive to tiny differences in their input
powers.
\end{abstract}

% We propose and study  stationary and oscillatory light propagation through a pair of nonlinear lossy optical waveguides 
% embedded  in an active medium which amplifies evanascent waves. Due to interference the resulting gain is dependent on the relative phase
% in the two waveguides.  
% Above a gain threshold we find stable light propagation in a wide range of parameters. We observe a cascade of different regimes
% supporting static, oscillating, and even chaotic propagation modes. Nontrivial light dynamics in these modes can be used for
% various optical devices like switches and comparators.

%The loss compensation is achieved in a finite range of gain coefficients so
%that the structure does not require any fine-tuning of its parameters.
%The system does not exhibit an uncontrollable growth of optical modes.

\maketitle

{\it Introduction.---} 
Nonlinear directional  couplers are important for various applications in integrated optics,
such as power-sensitive
switches and  polarization beam splitters.
The twin core coupler is based on the coherent light exchange between two optical waveguides
placed in close proximity.
For low input intensities, the full power oscillates periodically between the two waveguides;
for higher input levels, the total power is
 selftrapped mainly
  in one of the two channels \cite{NL_coupler}.

The performance of the device can be improved 
--- the switching power reduced while the length shortened --- by utilizing material losses.
The effect of absorption which is usually regarded as an unavoidable hinder, is therefore turned into advantage here.
 The dissipation is balanced by introducing  gain into one of the waveguides \cite{premaratne2011}.
Thus, 
a nonlinear
coupler composed
of one core with a certain amount of gain and another one
with an equal amount of loss switches the entire power to one waveguide 
\cite{CSP}.
Recently this type of a coupler has attracted a lot of attention as an experimental
realisation of a  $\mathcal{PT}$-symmetric system \cite{PT}. 
It is worth noting here 
 that the operation of the $\mathcal{PT}$-symmetric coupler requires the fine-tuning
 of gain and loss, to secure their 
    exact compensation.
% We also note that  the gain in this system is insensitive to the phase relation between the two waveguides.
 We also note that  the power switching is accompanied by the 
unbounded power growth in one of the arms of the 
device ---
the growth  not saturable  by nonlinearity \cite{PT}.

 In this Letter, we propose  a  conceptually new configuration of gain and loss in the 
 directional coupler. The arrangement consists of two lossy waveguides placed in an active medium. 
 Instead of providing power gain in the core of a waveguide, the  structure 
 boosts the evanescent
 fields which couple the two channels due to their close proximity.  
Choosing the symmetric gain configuration,  the energy is pumped into
the in-phase linear mode of the two-waveguide system while the anti-phase normal mode  remains lossy. 

The operation of the outlined device (the actively coupled pair of waveguides, or simply
 ``$\mathcal{AC}$-coupler")
is conditional on the presence of nonlinearity.
As the symmetric mode starts to grow, it activates the nonlinear response
in each waveguide; the nonlinearity couples the symmetric  to  antisymmetric mode
 which 
  drains the energy out of the system
securing the overall power balance.  The concept admits  generalisations  to networks of 
 waveguides, with various coupling geometries. It can also find applications outside the realm of optics, e.g. in the context of 
 structured metamaterials \cite{lazarides2013} and interacting exciton-polariton condensates \cite{aleiner2012}.

The $\mathcal{AC}$-coupler  is structurally stable --- in order to ensure the gain-loss balance, one need not
 strive to secure the
perfect equality of gain and loss.
Instead, the loss compensation is achieved in a finite band of gain coefficients.
Unlike the usual nonlinear optical coupler, 
and unlike the $\mathcal{PT}$-coupler,
the observable regimes
in  the new configuration are determined
by the system parameter values rather than initial conditions.
Furthermore, the system does not exhibit any uncontrollable growth of optical modes.

   \begin{figure}[t]
 \begin{center}
       \includegraphics*[width=0.9\linewidth]{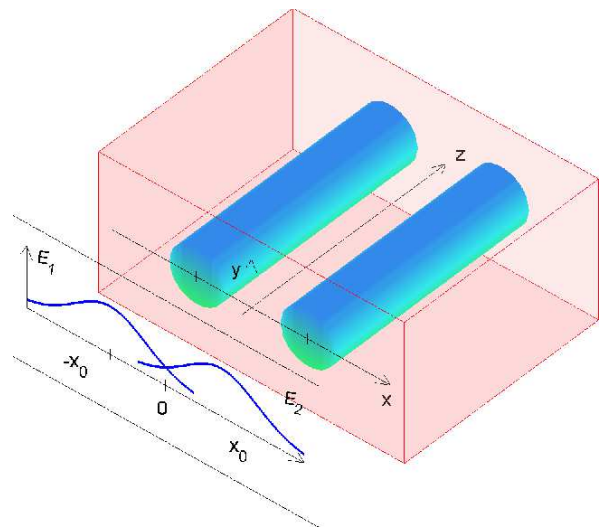}
            \caption{(Color online)   The $\mathcal{AC}$ coupler: two parallel lossy waveguides 
   coupled via an active medium. The stand-alone vertical plane shows 
   the eigenfunctions $E_1(x,0)$ and $E_2(x,0)$.   
             \label{Fig1}}
 \end{center}
 \end{figure}

{\it Model.---}
In a single-mode optical waveguide,
the optical field 
 is described by a complex amplitude $\Psi(x,y,z) = E(x,y) \psi(z)$, 
where $(x,y)$ is the plane transversal to the waveguide axis $z$.
The eigenfunction $E(x,y)$ decays away  from the waveguide core
%(where $x=y=0$)
and  can be chosen to be real and everywhere positive, with
the norm $\int E^2 dx \, dy = 1$. 
%In a lossy waveguide,
%$\psi(z)$ decays exponentially as $z \to \infty$.

 We consider two parallel identical waveguides. 
Denoting
$x$  the coordinate in the direction connecting their centres, we
choose the origin halfway between them. The individual 
eigenfunctions $E_1$ and $E_2$ are then centred at  $x= -x_0$
and $x=x_0$, respectively,  and satisfy 
$E_{1}(x,y) = E_2(-x,y)$.
The waveguides are embedded in the active medium (Fig.\ref{Fig1})
 and gain power through 
 the response of the  medium to 
 the  evanescent
part of their  fields, $\Psi_1$ and $\Psi_2$.

The integral gain of the amplitude  $\psi_n(z)$ 
over the entire active region is $\int \alpha E_n(x,y)  \Psi dx dy $,
where $\Psi(x,y,z)$ is the sum of the two evanescent fields:
$\Psi= \Psi_1(x,y,z)+\Psi_2(x,y,z)$.
The coefficient $\alpha(x,y) \geq 0$ characterises the active
properties of the medium. 
% Here, we assume a symmetric arrangement: $\alpha(x,y)=\alpha(-x,y)$.
The integral gain in each waveguide is then
$a_{n1}\psi_1+ a_{n2} \psi_2$,
where
\[
a_{nm} = \int  \alpha(x,y)
E_n(x,y) E_m(x,y) dx dy, \quad  n,m=1,2.
% \label{eq4}
\]

With the gain added,   the amplitudes $\psi_1$ and $\psi_2$ 
satisfy
\begin{subequations}
\label{B}
 \begin{align}
 \frac{d \psi_1}{dz} + {\Gamma}  \psi_1 = i {\mathcal{T}}  \psi_2 + i {\beta}   |\psi_1|^2\psi_1+   \sum a_{1n} \psi_n,  
\label{B2}
\\
 \frac{d \psi_2}{dz} + {\Gamma}  \psi_2 = i{\mathcal{T}} \psi_1 + i {\beta}  |\psi_2|^2\psi_2 + \sum a_{2n} \psi_n.
\label{B1}
\end{align}
\end{subequations}
%  In \eqref{B}, 
Here
$\Gamma$ is the loss rate,
$\beta$ the nonlinearity
strength, and
$\mathcal{T}$  quantifies light tunneling between the guides \cite{Snyder_Love}.
In what follows,  we assume the  focussing nonlinearity, $\beta > 0$.
We  scale $\psi_{1,2}$ so that $\beta=1$
and normalise $\mathcal{T}$ to 1.

We note that $a_{12}=a_{21}$. Assuming symmetric density distributions
$\alpha(x,y)=\alpha(-x,y)$ we also have
%In this letter we restrict ourselves to  symmetric density distributions $\alpha(x)=\alpha(-x)$;
%for these,
 $a_{11}=a_{22}$. 
 From $E_{1,2}(x,y)>0$ it follows that  $a_{mn}>0$.
 Using the 
Schwartz inequality, one readily checks that 
 $a_{11}$ is always greater than $a_{12}$.
 All coefficients become equal only when  the active region is very thin: $\alpha(x,y) = \alpha_0(y) \delta(x)$.
 In this case we have  
$a_{11}=a_{12}=\int \alpha_0(y)   E^2_1(0,y)  dy$.  % ($m,n=1,2$). 

In the general situation of 
 symmetrically  distributed gain, we introduce
the net loss rate $\gamma=\Gamma-a_{11}$ and the active coupling ($\mathcal{AC}$) coefficient $a \equiv a_{12}$ to obtain
\begin{subequations}
\label{B1}
\begin{align}
 \frac{d \psi_1}{dz} + \gamma  \psi_1 = i \psi_2 + i   |\psi_1|^2\psi_1+   a \psi_2,  
\label{B1-2}
\\
 \frac{d \psi_2}{dz} + \gamma  \psi_2 = i \psi_1 + i  |\psi_2|^2\psi_2 + a \psi_1.
\label{B1-1}
\end{align}
\end{subequations}
In this Letter,  we consider the regime where $\gamma > 0$.

\begin{figure}[t]
 \begin{center}
    \includegraphics*[width=0.99\linewidth]{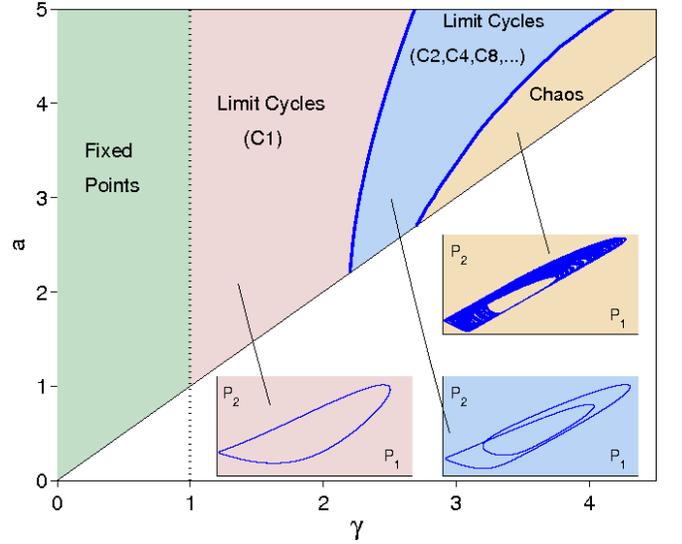}
           \caption{(Color online) The chart of attractors on the $(\gamma , a)$ plane.
           In the empty area below the $\gamma=a$ line,   %the coupler is trivially damped:
            all initial conditions decay to
           zero as $t \to \infty$.
            In  the green strip, the attractors are the fixed points $\mathrm{F}^\pm$.
           Pink marks the region where all trajectories wind onto one of the two limit cycles
           $\mathrm{C}^\pm$, with one simple oscillation per period (hence ``C1").
           In the blue domain  the attracting cycles have $2^n$
           (i.e. 2,4,8, ...)  simple oscillations per period.
           The ochre colours the area where the attractors are predominantly chaotic;
            these consist of repeated oscillations  with randomly selected amplitudes. 
           %  The chaotic area is interleaved with narrow strips where the 
           %   interleave with thin regions of limit cycles with odd numbers of oscillations per period. 
           Three  insets  are phase portraits on the $(P_1,P_2)$ plane:
          a cycle with one and two
            oscillations per period, and a chaotic attractor.
             \label{Fig2}}
 \end{center}
 \end{figure}

{\it Dynamics of coupled beams.---} 
When the net
 loss exceeds the $\mathcal{AC}$ gain
 ($\gamma>a$),
 all solutions of \eqref{B1} decay to zero:   % (irrespectively of the choice of  $\beta$):
\be
\frac{d P}{dz}  \leq   2(a-\gamma)P.
\label{E2}
\ee
Here $P=  |\psi_1|^2+ |\psi_2|^2$ is
the total  power of light in the coupler. 
Thus in the  region below the $\gamma=a$ line in Fig.\ref{Fig2}, 
 the origin $\psi_1=\psi_2=0$ is a globally stable fixed point.

  Introducing the symmetric and antisymmetric
  normal modes $u=\psi_1+\psi_2$ and $v =\psi_1-\psi_2$
diagonalises the linear part of \eqref{B1} --- but couples the nonlinear terms:
\begin{subequations}
\begin{align}
\frac{d u}{dz} +(\gamma -a-i) u =
\frac{i}{4}
[(|u|^2+ 2|v|^2)u +  v^2  u^*],
\label{p} \\
\frac{d v}{dz} + (\gamma +a+i) v=
\frac{i }{4}
[(2|u|^2+|v|^2)v + u^2 v^*].
\label{m}
\end{align} 
\end{subequations}
According to \eqref{p}, making $a$ greater than 
 $\gamma$ turns  the origin into an unstable fixed point.
 In the $a>\gamma$ regime with $v(0)=0$, the 
 symmetric mode grows without bound. However the blow-up of the 
 $u$-mode can be arrested by its nonlinear 
 coupling to the antisymmetric mode.

Decomposing the complex amplitudes as 
$\psi_1=\sqrt{P_1} e^{i \phi}$ and $\psi_2= \sqrt{P_2} e^{i (\phi+  \theta)}$, 
we observe that the common part of the phases of the two beams
is expressible through the powers they carry  and their phase difference:
\[
\phi(z)=  \int_0^z \left[ P_1+ \sqrt{ \frac{P_2}{P_1} } (\cos \theta+ a \sin \theta)             \right]     dz + \phi(0).
\]
% Therefore equations \eqref{B1-1} constitute a three- (rather than four-) dimensional dynamical system.
Therefore out of four real variables in the system  \eqref{B1} only 
% $P_{1,2}$  and $\theta$
three represent independent degrees of freedom.

%To make the three-dimensional dynamics more explicit, 
%In order to gain insight into the  dynamics of this system of the three mutually coupled variables,
% \eqref{B1-1} 
%with % the focussing nonlinearity and 
%$a > \gamma$,
%\eqref{B1}, 

This fact is made obvious by transforming 
 the system to an explicitly three-dimensional form:
\begin{subequations}
\label{E1} 
\begin{align}
{\dot X}=-\gamma X  - Y,   \label{C1}  \\
{\dot Y}=-\gamma Y +  X -  XZ,    \label{C2}   \\
{\dot Z}=-\gamma Z + a r + XY.    \label{C3}  
\end{align} 
\end{subequations} 
Here $X=\frac12(|\psi_1|^2-|\psi_2|^2)$ measures the power imbalance between the two waveguides;
 $Y=\frac{i}{2} (\psi_1\psi_2^*-\psi_1^*\psi_2)$
characterizes the energy flux from the first to the second channel, and
$2aZ$ --- where  $Z=
\frac12( \psi_1\psi_2^*+\psi_2\psi_1^*)$ ---
is the total gain in the system. The Stokes variables $X,Y$, and $Z$  are three components of the vector ${\bf r}$,  with  $r=\sqrt{{\bf r}^2}=P/2$.
The overdot indicates differentiation with respect to the fictitious time variable,
$t=2z$, which we introduce for convenience of analysis.

The beam powers $P_{1,2}$ and
the phase difference $\theta$ can be easily reconstructed from the Stokes variables:
$P_{1,2} = r \pm X$ and $\tan \theta= Y/Z$.
% The common part of the phases   $\phi= \frac12 \int   \left[ (r+X)^2 + Z+aY \right] \frac{dt}{r+X}$.
 We also note a similarity between equations \eqref{E1} and the Lorentz system \cite{Sparrow}.
 % Like the Lorentz system, Eqs.\eqref{E1} are quadratic and symmetric  under the reflection $X \to -X$, $Y \to -Y$.

\begin{figure}[t]
 \begin{center}
      \includegraphics*[width=\linewidth]{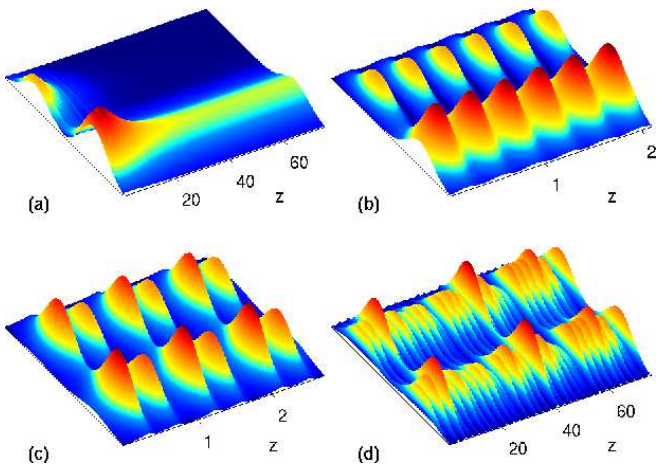}
                                    \caption{(Color online) Characteristic regimes of the 
                 $\mathcal{AC}$ coupler.
         Shown is $|E_1(x,0) \psi_1(z) + E_2(x,0) \psi_2(z)|^2$, with
         the eigenfunctions exemplified 
         by 
         $E_{1,2}(x,0)=\frac{1}{\sqrt{2\pi}\nu} \exp \left( \frac{-(x \pm x_0)^2}{2\nu^2} \right)$,  $\nu = \frac35 x_0$. 
 %        (These eigenfunctions correspond to  graded index-type waveguides.) 
          (a) The stationary regime $\mathrm{F}^+$ evolving out of the 
          initial condition with $P_1=30$, $P_2=33$, $\theta=0$.
           Here $a=1$, $\gamma=0.1$.
           (b) The limit cycle $\mathrm{C}^+$ $(a=9$, $\gamma=1.5$).
           (c) The limit cycle with two oscillations per period 
          ($a=9$, $\gamma=4.2$). 
         (d) The chaotic attractor ($a=9$, $\gamma=7.6$). 
                  }
             \label{Fig3}
 \end{center}
 \end{figure}

{\it Symmetry-broken  fixed points.---}   Like the Lorentz system, 
Eqs.\eqref{E1} are  symmetric  under the reflection of $X$ and $Y$.
% $X \to -X$, $Y \to -Y$.   The two main ingredients of the dynamics are  fixed points and limit cycles.
%Indeed, like the Lorentz case, the system \eqref{E1} has  two  fixed points
%besides the one at the origin.
As the $\mathcal{AC}$ coefficient is increased beyond $a=\gamma$
in the weakly-dissipative regime ($\gamma < 1$), the fixed point at the origin suffers a symmetry-breaking 
 (pitchfork) bifurcation. Two stable fixed points $\mathrm{F}^+$ and $\mathrm{F}^-$ are born, supercritically.
These points share the values of $r$ and $Z$,
\be
r=
(\gamma^2+1)
  \frac{a}{\gamma},
  \quad
  Z= \frac{\gamma}{a} r, 
\label{C11}
\ee
but have opposite  $X$ and $Y$:
\be
X=  - \frac{\sigma}{a}  r, \quad
Y=     \sigma                   \frac{\gamma}{a} r.
\ee
The point F$^+$ has  $\sigma>0$ and F$^-$ has  $\sigma<0$,
where
$\sigma^2= (a^2 - \gamma^2)/(\gamma^2+ 1)$.
The two points are therefore mirror images of each other.

\begin{figure}[t]
 \begin{center}
    \includegraphics*[width=\linewidth]{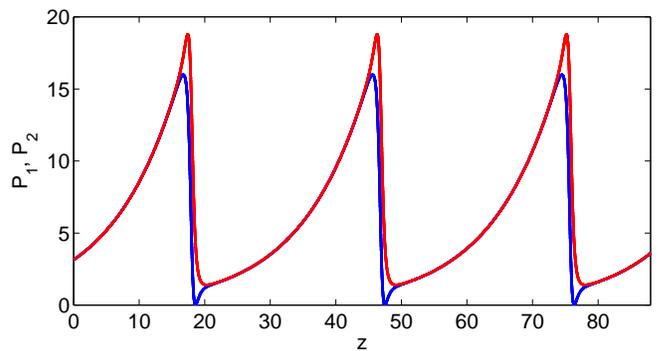}
           \caption{(Color online) The integrate-and-fire limit cycle
           arising for $\gamma$ close to $a$. (In this plot, $\gamma=1.9$
           and $a=2$.)
           The red 
           and blue line show $P_2$ and $P_1$, respectively.
          }
             \label{Fig4}
 \end{center}
 \end{figure}

Either of the two fixed points   represents a pair of light beams with constant, $z$-independent, power in each channel.
The point F$^+$  corresponds to a greater power in the second channel ($P_1/P_2 \approx \gamma^2/2$
for small $\gamma$), while its mirror reflection F$^-$ has the inverse power ratio.
The phases $\phi_1$ and $\phi_2$ pertaining to the points F$^\pm$ are linear functions of $z$,
with the phase difference $\theta$ remaining constant.

 Fig.\ref{Fig3}(a) depicts the evolution of the input with
 a small power imbalance. We observe that 
 taking $P_2(0)$ only 
 slightly greater than $P_1(0)$ is sufficient to select 
 the stationary regime ($\mathrm{F}^+$) with $P_2 \gg P_1$.

{\it Periodic and chaotic regimes.---} 
Linearizing Eqs.\eqref{E1} about  either of 
the fixed points $\mathrm{F}^\pm$ yields the Jacobian matrix with
one real negative and
two complex-conjugate  eigenvalues $\lambda$.
Assume the 
 active coupling is fixed above the threshold value $a = 1$
 and the net loss $\gamma$ is varied.
As  $\gamma$ is increased through  $\gamma_c=1$, 
the complex eigenvalues cross  from $\mathrm{Re} \, \lambda<0$ to $\mathrm{Re} \, \lambda>0$.
This is the signature of the Hopf bifurcation where both $\mathrm{F}^+$ and $\mathrm{F}^-$ lose their stability
while
two stable limit cycles of small radius
are born --- one around each fixed point.  % \cite{marsden1976}. 

The limit cycles describe 
periodic variation of the powers $P_{1,2}$ carried by the two waveguides [Fig.\ref{Fig3}(b)].
We denote $\mathrm{C}^+$ the cycle born around the point $\mathrm{F}^+$ and $\mathrm{C}^-$ the cycle 
bifurcating from $\mathrm{F}^-$.

Like their parent fixed points $\mathrm{F}^\pm$, the limit cycles $\mathrm{C}^\pm$ are symmetry broken.
 The cycle $\mathrm{C}^+$ has the second waveguide carrying higher  power than the first one, while its
 mirror-reflected counterpart
 $\mathrm{C}^-$ has $P_1(z)>P_2(z)$ for all $z$.
When $\gamma=\gamma_c$, the  frequency of each cycle is given by $\omega= \mathrm{Im} \, \lambda =\sqrt{a^2-1}$.
As $\gamma$ is increased beyond $\gamma_c$, the radius grows and the frequency changes.

The limit cycles  with $\gamma$  near $a$
 display an integrate-and-fire switch dynamics
(Fig.\ref{Fig4}). Namely, 
 the powers  in the two waveguides
grow slowly and  synchronously, with
 one of these (say, $P_1$) remaining only slightly smaller than the
 other power ($P_2$). This is followed by a quick discharge, where the 
 first waveguide loses practically all its power ($P_1 =0$) while 
 $P_2$ remains nonzero.

\begin{figure}[!t]
 \begin{center}
    \includegraphics*[width=\linewidth]{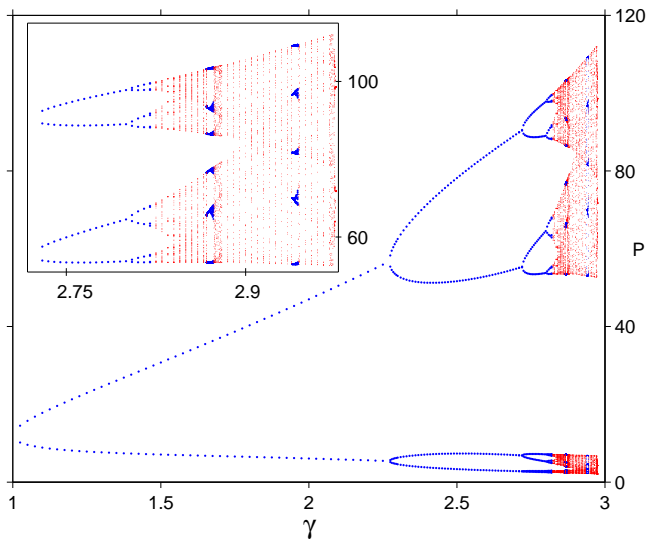}
{ \caption{(Color online) The period-doubling transition to chaos for $a=3$. 
The local minima and maxima of the function $P(z)$ are plotted against the 
corresponding value of the loss coefficient.
The inset shows a fine structure of the chaotic region.
Periodic attractors are marked by blue; chaotic by red.
}
                  \label{Feigenbaum}}
 \end{center}
 \end{figure}

As the loss $\gamma$ is increased with
 the value of
active coupling  fixed above $a=2.2$,
the limit cycle suffers a period doubling. 
The emerging periodic attractor with 
two oscillations per period is shown  in Fig.\ref{Fig3}(c)
and the middle  inset in Fig.\ref{Fig2}. 
Increasing $\gamma$  further, the limit cycle 
undergoes a cascade of higher 
period-doubling bifurcations (Fig.\ref{Feigenbaum}). 
This
culminates in  the emergence of chaotic attractors 
[Fig.\ref{Fig3}(d) and the top inset in Fig.\ref{Fig2}]. 
The set of parameter values corresponding to periodic attractors
with $2^n$ oscillations per period
($n=1,2,...$), is marked blue   in Fig.\ref{Fig2},
and the ``chaotic domain" is ochre-colored.

{\it Conclusions.---}  We conclude our
study of the
$\mathcal{AC}$ coupler by
 crystallising   its key difference from the
$\mathcal{PT}$ symmetric device.
While the latter balances gain in one waveguide with loss in the other,
the former  treats its two channels equally. 
It is the symmetric and antisymmetric normal modes 
(rather than the two guides themselves) that serve as the 
gain and loss agents in the $\mathcal{AC}$ configuration.

An important advantage of the $\mathcal{AC}$ coupler is its structural stability.
For the given loss rate, the system supports stationary and periodic 
light beams in a wide range of gain coefficients rather than for a 
particular value of $a$. This is a fundamental distinction from the 
$\mathcal{PT}$-symmetric coupler where one has to tune the gain to
match the loss exactly \cite{PT}.

The new properties exhibited by the device can be utilised in
a variety of applications. One possible setting is an optical analog of the 
voltage comparator which  swings its output to one of two values depending on the
relation between its two input voltages. The  $\mathcal{AC}$ coupler can operate 
either in the stationary regime, where
the two output values are given by the 
fixed points of the system \eqref{E1}, or as an integrate-and-fire switch
(Fig.\ref{Fig4}),
where one of the waveguides is left powerless periodically. The period of this oscillator
can be tuned simply by varying the distance between the waveguides.
Unlike the $\mathcal{PT}$-symmetric coupler, no input can trigger an 
uncontrollable growth of optical modes in the 
 $\mathcal{AC}$-switch.

{\it Acknowledgments.---} N.A. and I.B.'s sabbatical leave in Auckland was funded via
visiting
fellowships of the New Zealand Institute for Advanced Study.
The project was
also supported by the NRF of South
Africa (Grants UID 85751 and 78950) and the Vera Davie study and research bursary from UCT.

\end{document}